\documentclass[12pt,preprint]{aastex}

\def\gta{\ifmmode {\mathbin{\lower 3pt\hbox   
    {$\,\rlap{\raise 5pt\hbox{$\char'076$}}\mathchar"7218\,$}}}
    \else {${\mathbin{\lower 3pt\hbox
    {$\rlap{\raise 5pt\hbox{$\char'076$}}\mathchar"7218\,$}}}
    $}\fi}
\def\lta{\ifmmode {\,\mathbin{\lower 3pt\hbox   
    {$\,\rlap{\raise 5pt\hbox{$\char'074$}}\mathchar"7218\,$}}}
    \else {${\mathbin{\lower 3pt\hbox
    {$\rlap{\raise 5pt\hbox{$\char'074$}}\mathchar"7218\,$}}}
    $}\fi}

\shorttitle {Front stalling during a non-PRE double-peaked burst}
\shortauthors {Bhattacharyya and Strohmayer}

\begin{document}

\title {Signature of Temporary Burning Front Stalling from a Non--Photospheric Radius 
Expansion Double-peaked Burst}

\author {Sudip Bhattacharyya\altaffilmark{1,2}, and Tod
E. Strohmayer\altaffilmark{2}}

\altaffiltext{1}{Department of Astronomy, University of Maryland at
College Park, College Park, MD 20742-2421}

\altaffiltext{2}{X-ray Astrophysics Lab,
Exploration of the Universe Division,
NASA's Goddard Space Flight Center,
Greenbelt, MD 20771; sudip@milkyway.gsfc.nasa.gov,
stroh@clarence.gsfc.nasa.gov}

\begin{abstract}

Non-photospheric-radius-expansion (non-PRE) double-peaked bursts may
be explained in terms of spreading (and temporary stalling) of
thermonuclear flames on the neutron star surface, as we argued in a
previous study of a burst assuming polar ignition.  Here we analyze
Rossi X-ray Timing Explorer (RXTE) Proportional Counter Array (PCA)
data of such a burst (but with a considerably different intensity
profile from the previous one) from the low mass X-ray binary (LMXB)
system 4U 1636--536, and show that this model can qualitatively
explain the observed burst profile and spectral evolution, if we
assume an off-polar, but high-latitude ignition, and burning front
stalling at a higher latitude compared to that for the previous
burst. The off-polar ignition can account for the millisecond period
brightness oscillations detected from this burst. This is the first
time oscillations have been seen from such a burst. Our model can
qualitatively explain the oscillation amplitude measured during the
first (weaker) peak, and the absence of oscillations during the second
peak. The higher latitude front stalling facilitates the first clear
detection of a signature of this stalling, which is the primary result
of this work, and may be useful for understanding thermonuclear flame
spreading on neutron stars.

\end{abstract}

\keywords{accretion, accretion disks --- relativity --- stars: neutron ---
X-rays: binaries --- X-rays: bursts ---  X-rays: individual (4U 1636--536)}

\section {Introduction} \label{sec: 1}

Thermonuclear burning of matter accumulated on the surfaces of
accreting neutron stars produces type I X-ray bursts (Woosley, \& Taam 1976;
Lamb, \& Lamb 1978).  Although the light curves of most of these
bursts have single-peaked structures, some of the strongest bursts
from a given source show double-peaked profiles in higher energy
bands. These can be explained in terms of photospheric radius
expansion (PRE; due to radiation pressure), and subsequent contraction
(Paczynski 1983).
However, double-peaked structures are also seen in weak non-PRE bursts, and in
such cases, these structures appear both in low and high energy bands.
Such bursts have so far only been observed infrequently from a few
sources (Sztajno et al. 1985). Until recently,
models for these bursts did not simultaneously explain their intensity
profiles, spectral evolution, X-ray emission area increase, and rarity
(Fujimoto et al. 1988; Fisker, Thielemann, \& Wiescher 2004; Regev, \&
Livio 1984; Melia, \& Zylstra 1992). However, the recent model of
Bhattacharyya \& Strohmayer (2006), based on thermonuclear flame
spreading on neutron stars, can qualitatively explain these
properties. According to this model, such bursts are ignited close to
a rotational pole (which explains the lack of millisecond period brightness
oscillations), and then the burning front propagates (as a more or
less $\phi$-symmetric belt; $\phi$ is the azimuthal angle) towards the
equator. As the front approaches the equator, it stalls for a
few seconds before speeding up again into the opposite hemisphere. The
stalling of the front allows the burning region to cool, and the
luminosity to decrease.  After a few seconds, as the flame starts
spreading again, the increased temperature and emission area cause the luminosity to
increase again, and hence a double-peaked structure appears in the
burst profile. The physical reason for the temporary stalling of the
front is still uncertain, although it might be caused by the
accretion-induced pole-ward motion of burning shell matter
(Bhattacharyya \& Strohmayer 2006).

Our proposed model suggests that the non-PRE double-peaked bursts can
be a useful tool to understand thermonuclear flame spreading on
neutron star surfaces. Such an understanding may be important to
constrain stellar surface parameters, as well as to model burst
oscillations and their frequency evolution during burst rise
(Bhattacharyya \& Strohmayer 2005a; 2005b). Accurate modeling of burst
oscillations has importance for constraining neutron star masses and
radii, and hence the dense matter equation of state
(Bhattacharyya et al. 2005).  In this Letter, analysing RXTE PCA data
of a non-PRE double-peaked burst (with significantly different
intensity profile from the burst analysed in Bhattacharyya \&
Strohmayer 2006) from the LMXB 4U 1636--536, (1) we report the
discovery of oscillations for the first time from a non-PRE
double-peaked burst, (2) we show that these oscillations can be
accommodated in the spreading model, when high latitude but off-polar
ignition is assumed (which still explains the rarity of these bursts),
and most importantly, (3) we find a strong signature of temporary
burning front stalling during this burst (which has been lacking in
case of the Bhattacharyya \& Strohmayer 2006 burst). We argue that
these new results provide additional support for the idea that
thermonuclear flame spreading can account for many of the features of
non-PRE double-peaked bursts.

\section {Data Analysis and Model Calculations} \label{sec: 2}

We analyze the RXTE PCA archival data of a double-peaked burst (date
of observation: Feb 28, 2002; ObsId: 60032-05-15-00) from 4U
1636--536.  The second peak height $(\sim 5000$ counts/s/PCU)
is about three times larger than the first peak height
(Fig. 1). This burst is weaker than PRE bursts observed from this
source (Strohmayer et
al. 1998). The similarity of the burst profile in different energy
bands also indicates that this is not a PRE burst.
We discovered oscillations (near the stellar spin
frequency $\sim 582$~Hz; Strohmayer \& Markwardt 2002) during the
first intensity peak, but not during the second. We calculated the
power spectrum of 1 s of data during the first peak
(Fig. 1). At 2 Hz resolution, we find a peak power of 13.4 at the
frequency $\sim 581$~Hz. The probability of obtaining a power this
high in a single trial from the expected $\chi^2$ noise distribution
(4 dof) is $\approx 2.18 \times 10^{-5}$. As the oscillations have
been searched at the known frequency (which does not evolve by 
more than $\approx 6$ Hz; Giles et al. 2002), 
multiplying by the number (=3) of trials, we get a significance of
$6.55 \times 10^{-5}$, which implies a $\sim 4\sigma$ detection.  We
also fit the phase-folded light curve of 1 s of data (marked with the
vertical lines in Fig. 1) with a combined model of a sinusoid
and a constant ($\chi^2/{\rm dof} = 10.67/13$, as opposed to 
$\chi^2/{\rm dof} = 25.13/15$ for a constant model) and find that the 
oscillation rms amplitude is $0.082\pm0.022$. 

In order to track the burst spectral evolution, we break
the burst profile into smaller time bins, and for each bin perform
spectral fitting using a single temperature blackbody model (bbodyrad
in XSPEC), as generally burst spectra are well fit by a blackbody.
During the fitting, we fix the
hydrogen column density $N_{\rm H}$ at a value
$0.56\times10^{22}$~cm$^{-2}$ (Bhattacharyya \& Strohmayer 2006).  The
fitting is performed after subtracting the persistent (i.e., preburst)
emission from the emission during the burst.  We show the time
evolution of the best fit values of temperature and radius (calculated
from the bbodyrad ``normalization'' parameter) in Fig. 2. 
This figure shows that these parameters are well
correlated with the intensity profile, and the
radius, which is a measure of the source emitting area, increases with
time in a manner indicative of flame spreading. Here
we note that some of the properties of this burst were also reported
by Jonker et al. (2004), which they used only to conclude that this was not
a PRE burst. The reduced $\chi^2$ values
for some of our time bins are high $(> 1.5$ for two out of 14
bins). We previously argued that this might result from temperature
variations in the burning region at a given instant.
The persistent emission may vary during the burst, which can have some
effect on the inferred radius values derived from spectral fitting. To
explore this possibility, we fit the background subtracted preburst emission with a 
model and then fit the background subtracted burst spectra with 
the best fit values (but varying the normalization) of this same model,
plus a blackbody. The resulting temporal evolutions of blackbody temperature and
radius were found to be very similar to those shown in Fig. 2,
providing additional confidence in the evolutionary tracks shown in this
figure.

We explain the observed data using a model (involving flame spreading and its
temporary stalling), which is very similar to that
of Bhattacharyya \& Strohmayer (2006; see \S~1), except here we consider 
ignition at high latitudes (instead of polar ignition). This off-polar 
ignition gives rise to an expanding $\phi$-asymmetric hot spot, which causes the oscillations
during the first intensity peak. At a certain $\theta$-location $(\theta$ is 
the polar angle), the $\theta$-ward motion of the flame
stalls temporarily, and its $\phi$-ward motion quickly gives the burning
region the shape of a $\phi$-symmetric belt (which causes the
disappearance of oscillations consistent with the observation), which then
expands $\phi$-symmetrically. The temporary stalling of the
$\theta$-ward motion, but not of the $\phi$-ward motion, is consistent
with the suggestion (Bhattacharyya \& Strohmayer 2006) that
this stalling is caused by accretion-induced pole-ward motion of
burning shell matter, while the accretion is conducted via a disk, and
hence is $\phi$-symmetric.
However, we model the observed burst intensity profile using a 
simplification, that the burst is ignited 
in a narrow $\phi$-symmetric belt. 
This does not cause any error (within the
context of our model) in the intensity profile after the maximum
of the first peak (as at this time, stalling makes the hot spot
$\phi$-symmetric), and before this time, this simplified model approximately
reproduces the intensity increase.
We make this simplification for the following reasons. (1) The low
signal to noise ratio data do not show any significant evolution of
the oscillation amplitude or frequency. Therefore, an understanding of
the burning region geometry evolution has to come from theoretical
calculations. (2) A rigorous theoretical calculation of thermonuclear
flame spreading considering all the main physical effects (e.g., that
of magnetic field) is not available at the present time (but see
Spitkovsky, Levin, \& Ushomirsky 2002). 
However, we note that in reality there must be a $\phi$-asymmetric
burning region during the first peak rise in order to explain
the oscillations. But this discrepancy will not be serious {\it as
long as} the actual $\phi$-asymmetric burning region can also
approximately reproduce the first intensity peak.
We examine this point now. From our model (Fig. 3), 
we find that at the time of the first
intensity peak, the $\phi$-symmetric burning region covers the
$\theta$-space from $\theta = 0^{\rm o}$ to $\theta = 40^{\rm o}$.
A belt this wide, but with a $\phi$-width of $\approx
260^{\rm o}$, produces an oscillation amplitude $0.072\pm0.011$
(consistent with the observed value), and an intensity that is 
within 9\% of the maximum of the first intensity peak. Therefore,
our simplified model can semi-quantitatively reproduce both
the observed intensity and the oscillation amplitude simultaneously.
Here we note that our assumed $\phi$-elongated burning region is not unrealistic,
as in this region the hot layers will puff up and ``slip'' in the
$\phi$-direction with respect to the stellar surface (in order to
conserve angular momentum; Cumming \& Bildsten 2000), 
causing a quicker spread of the flame in
this direction (compared to the $\theta$-direction) at a given latitude.
We also note that as this effect affects only the $\phi$-direction
speed of the flame, our usage of Spitkovsy et al.'s (2002) ageostrophic
flame speed (that does not include this effect) for $\theta$-ward
motion (see the next paragraph) is justified.
However, this $\phi$-elongated burning region geometry is only suggestive,
and a more realistic
geometry has to come from (statistically) better data and rigorous
theoretical studies. 

As mentioned above, we calculate the intensity profile
considering that the burst is ignited at a polar angle $\theta =
\theta_c(= 10^{\rm o})$ and a narrow $\phi$-symmetric belt expands in both
directions with an angular speed $\dot\theta(\theta) = F(\theta)$,
where $F(\theta) = 1/(t_{\rm total}\times\cos\theta)$ for $\theta \le
90^{\rm o}$ and $F(\theta) = 1/(t_{\rm total}\times\cos(180^{\rm
o}-\theta))$ for $\theta \ge 90^{\rm o}$.  Here $t_{\rm total}$ is the
front propagation time (from a pole to the equator) without any stalling.
Note that this particular expression of $F(\theta)$ (adopted from Spitkovsky et al.
2002; for weak turbulent viscosity) is {\it not} crucial for our qualitative results.
We assume that the stalling of the front happens between the polar angles $\theta_1$
and $\theta_2$ (with $\theta_1 < \theta_2 < 90^{\rm o})$:
$\dot\theta(\theta)$ decreases linearly from $\theta = \theta_1$ to
$\theta = \theta_m$, reaching a value $s/t_{\rm total}$, and then
increases linearly up to $\theta = \theta_2$ reaching a value
$F(\theta_2)$.  Between $\theta = \theta_2$ and $\theta = 180^{\rm
o}$, we assume $\dot\theta(\theta) = F(\theta)$.  To calculate the
temperature of a given location at a certain time, we assume that
after ignition of the fuel at that location, the temperature increases
from $T_{\rm low}$ to $(T_{\rm low} + (0.99\times (T_{\rm high} -
T_{\rm low})))$ following the equation $T(t) = T_{\rm low} + (T_{\rm
high} - T_{\rm low})\times(1-\exp(-t/t_{\rm rise}))$, and then decays
exponentially with an e-folding time $t_{\rm decay}$.  We compute
light curves and spectra using this model (see Bhattacharyya \&
Strohmayer 2006 for details), considering
Doppler, special relativistic, and general relativistic (gravitational
redshift and light-bending in Schwarzschild spacetime) effects.
In Fig. 3, for an example set of source parameter values, we show the
evolutionary tracks of model intensity and spectral parameters
(blackbody temperature and radius) that are qualitatively similar to
the observed features of the double-peaked burst (Fig. 2).  For
example, the model reproduces the general shape of the observed
intensity profile (including the $\sim 1/3$ ratio of first peak height 
to second peak height). For both data and
model, the blackbody temperature starts from a high value, decreases
up to the time when the intensity becomes minimum between the two
peaks, increases up to the time when the intensity reaches the second
peak, and declines after that. The inferred radius for both data and
model starts from a low value, increases up to the time when the
intensity reaches the first peak, remains almost unchanged up to the
time when the intensity becomes minimum between the two peaks,
increases again up to the time when the intensity reaches the second
peak, and remains almost unchanged after that. The almost constant
value of the inferred model radius during the decline of the first
intensity peak (marked with the two vertical lines in Fig. 3) is due to
the temporary stalling of the burning front. The occurrence of the
same feature during the observed burst (see Fig. 2) is therefore a
clear indication of front stalling.  
Note that for the emission from a localized hot spot, due to the lack 
of knowledge about the size and location of the spot, and the amount
of light-bending (which depends on stellar parameters), we cannot clearly 
interpret the inferred radius apart from suggesting that its square
is approximately proportional to the burning region area. Therefore, we do
not conclude anything directly from these inferred (from data) radii,
but compute similar radii (using stellar and hot spot parameters as
variables) from the model, and then compare between these two sets.
As we find the evidence of temporary front stalling from this 
comparison, we consider it to be strong.
We note that the more gradual rise of the model burst peaks
may be because of the delay between
ignition at depth and emergence of the radiation (Bhattacharyya \&
Strohmayer 2006), and the simplification (i.e., a $\phi$-symmetric
ignition) of the model.  
The observed higher second peak temperature compared to that of the
first peak (Fig. 2; which does not happen for the model)
may be due to the slightly higher burst maximum temperature $(T_{\rm high})$
during the second peak than that just after the burst onset (while 
we do not consider such a $T_{\rm high}$-variation in the model).
Such a temperature difference is not unrealistic, as for
accretion via a disk, the lighter element hydrogen (as opposed to helium)
is expected to spread more towards the pole. This would
cause a higher hydrogen-to-helium ratio near the pole (than near the
equator), and hence lower $T_{\rm high}$ value just after the burst
onset (as the burst ignites at a high latitude).  We also note that
the model burst duration scales with the parameter
$t_{\rm total}$ and hence can be adjusted by changing the value of
this parameter.

\section {Discussion and Conclusions} \label{sec: 3}

In this Letter we show that our simple model, based on thermonuclear
flame spreading and its temporary stalling, can qualitatively
reproduce the intensity and spectral profiles of a non-PRE
double-peaked burst from 4U 1636-53. The burst studied here is
different from the one described in a previous paper (Bhattacharyya \&
Strohmayer 2006) 
in the sense that the peaks of this burst 
have considerably different fluxes, while they were almost equal for the
previous one. The observed  radius profile of this burst is also different
from that of the previous burst (Fig. 2).
Moreover, this burst provides an additional challege to the
flame spreading model in that one must account 
for the detection of oscillations in the first peak.
In addition, although the general support for flame spreading
and its temporary stalling comes from this burst, as well as from the
previous burst, a clear (disentangled
from other effects) evidence of front stalling has been lacking
until this study, which we describe below.
During the stalling, the inferred
radius is not expected to change much (as the emission area does not), 
and this constancy in radius
value should correlate with the decrease of the intensity
and the temperature caused by the stalling (see \S~1). 
Therefore, this constancy in between two clear increases (due to 
flame spreading) of radius
values can give a very strong evidence of stalling (disentangled from other
effects). Such a clear evidence was missing from the previous burst. This 
is because, for the previous burst, we had to consider a low latitude stalling
(to reproduce the equal intensity peak heights), which made
the constancy in radius due to stalling confused with the radius constancy 
in the later stage of spreading, as these two were almost connected
(and the radius values were similar within error bars; 
Fig. 1 of Bhattacharyya \& Strohmayer 2006). This later radius constancy
happened when the burning region
expanded into the opposite hemisphere, and the front went out of the
sight (and hence the visible source area did not increase much). 
But for the burst studied here, this evidence is present
because of high latitude stalling (which is required to reproduce
the observed asymmetric intensity profile). Therefore,
this burst provides the first clearly disentangled signature of front stalling
(see the pairs of vertical lines in Figs. 2 \& 3).
Our model can also explain the rarity of these
bursts, as high-latitude ignition should be rare according to
Spitkovsky et al. (2002; Bhattacharyya \& Strohmayer 2006;
Bhattacharyya et al. 2000). Moreover, our model prefers the slower
front speed (for weak surface turbulent viscosity) of the two
front propagation speed scenarios presented by Spitkovsky et al. (2002).

Our discovery of millisecond period brightness oscillations from the
observed burst is the first such detection from a non-PRE
double-peaked burst, and our model can qualitatively account for the
observed rms amplitude, as well as the occurrence of these
oscillations only during the first intensity peak. Our model also
suggests that the details of the evolution of the burning region
geometry can be obtained from the 
data (with better signal-to-noise ratio) that show significant
oscillation amplitude and frequency evolution.
However, we note that although the observed oscillations
are consistent with our model involving thermonuclear flame spreading,
the possibility that these oscillations have the same origin as that
of burst tail oscillations can not be completely excluded. Therefore,
to more firmly establish our model, as well as to understand the
physics behind front stalling, it is essential to conduct rigorous
fluid dynamical calculations and simulations, and to expand the sample of 
non-PRE double-peaked bursts by observing 4U 1636--536 and
other similar sources for longer time periods with RXTE PCA and future
larger area detectors.

\acknowledgments

We thank the referee for pointing out a discrepancy and
helping to improve this Letter. 


{}

\clearpage
\begin{figure}
\hspace{-1.6 cm}
\epsscale{1.0}
\plotone{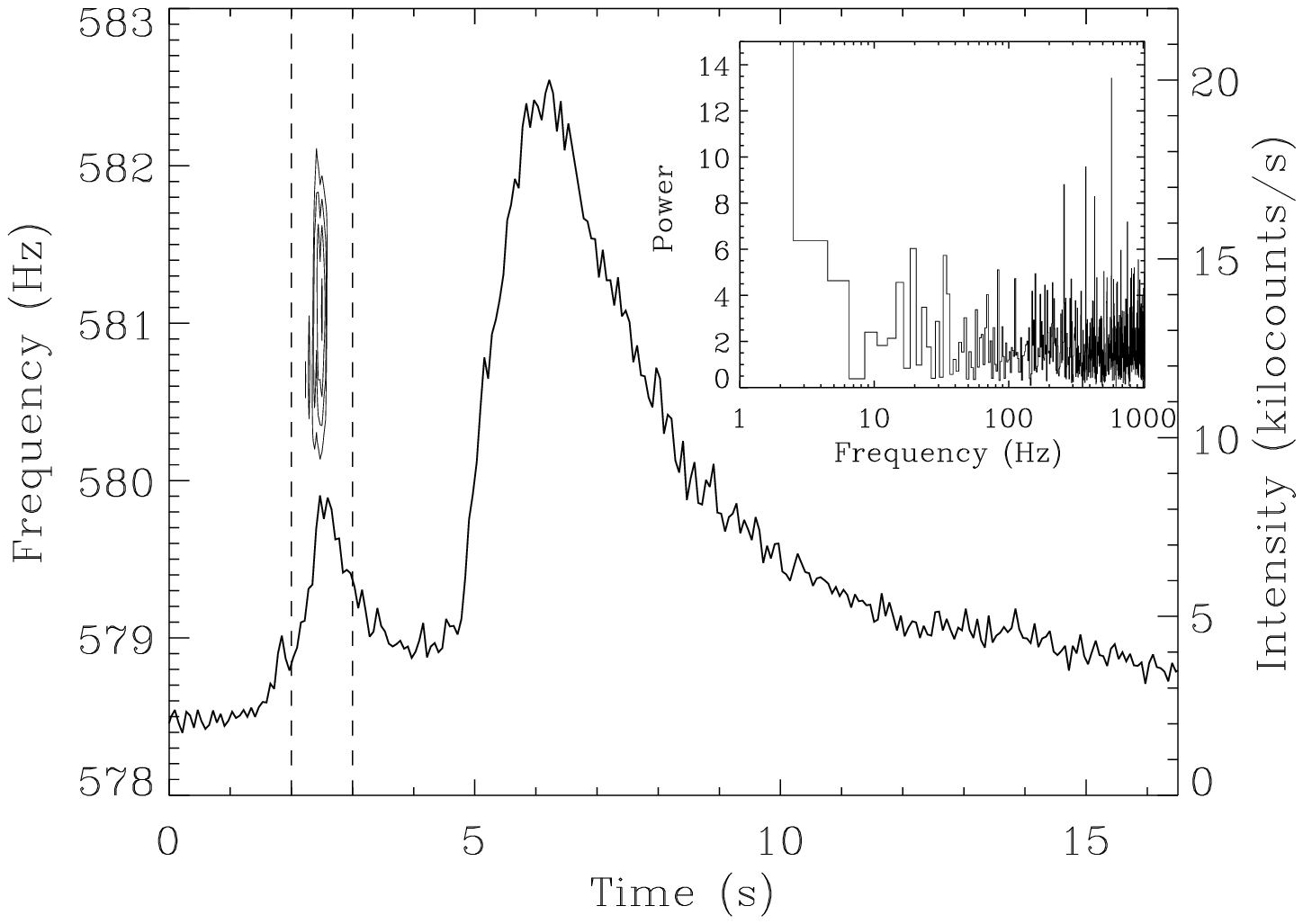}
\vspace{-6 cm}
\caption {Non-PRE double-peaked burst from 4U 1636--536.  The main
panel shows the PCA count rate profile (four PCUs on). The inset panel
shows the power spectrum of the 1 s interval (marked with vertical
dashed lines in the main panel) during the first peak, rebinned to 2
Hz resolution. The peak near 581 Hz implies a significant signal
power. Power contours using the dynamic power spectra (for 0.5 s
duration at 0.03 s intervals) are shown in the main panel (Strohmayer
\& Markwardt 1999).  Contours at power levels of 15, 18, 21, and 25 are
shown. These power contours show that oscillations occur only during
the first intensity peak.}
\end{figure}

\clearpage
\begin{figure}
\hspace{-1.6 cm}
\epsscale{1.0}
\plotone{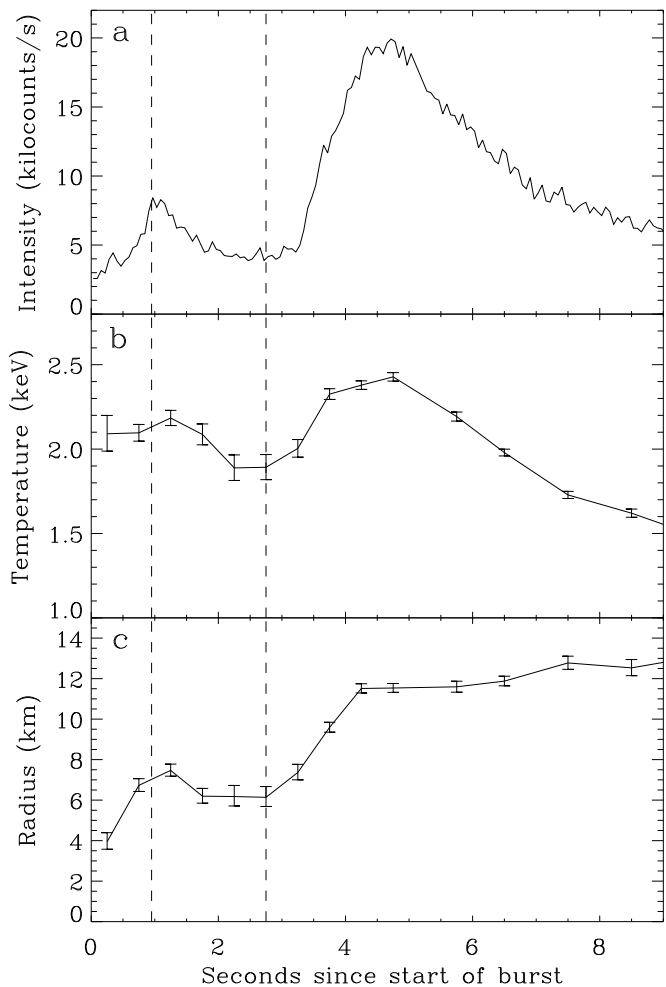}
\vspace{-4 cm}
\caption {Non-PRE double-peaked burst from 4U 1636--536.  Panel {\it
a} gives the bolometric burst profile. Panels {\it b} and {\it c} show
the time evolution of the blackbody temperature and the apparent
radius (assuming 10 kpc source distance) of the emission area
respectively, obtained by fitting the burst spectrum (persistent
emission subtracted) with a single temperature blackbody model. The
error bars are $1\sigma$.  Vertical dashed lines give the time
interval in which the radius (and hence the source emission area) does
not change, and the possible burning front stalling occurs.}
\end{figure}

\clearpage
\begin{figure}
\hspace{-1.6 cm}
\epsscale{1.0}
\plotone{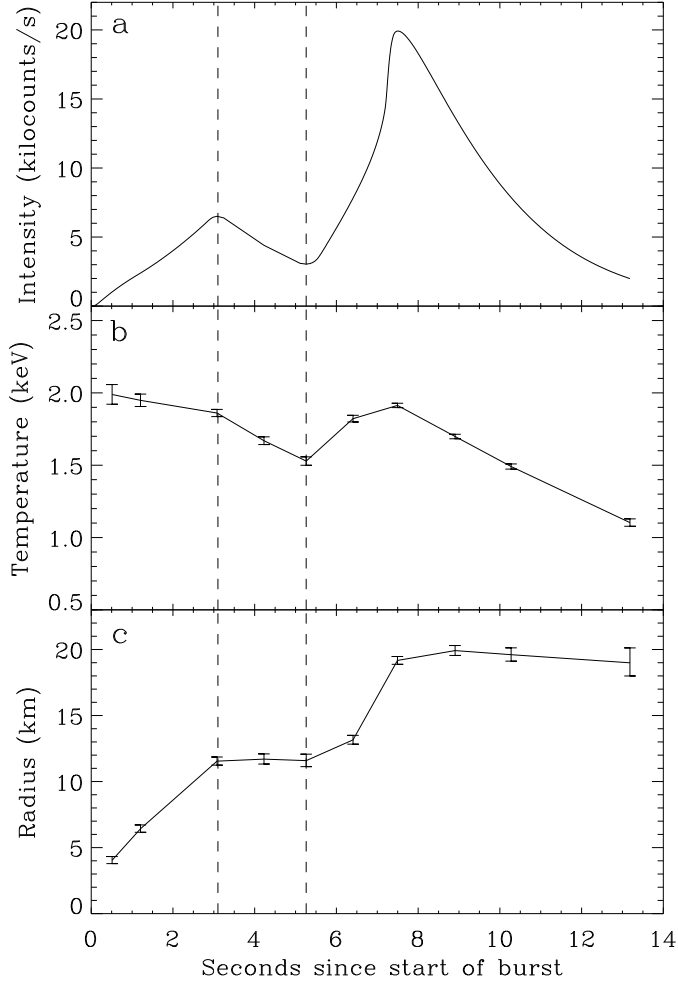}
\vspace{-4.5 cm}
\caption {Model (convolved with a PCA response matrix) of the
double-peaked burst: for all the panels, the burst is normalised so
that its second intensity peak has the same count rate as that of the
second peak of the observed burst. Panels are similar to those of
Fig. 2. Model parameter values are the following: stellar mass $M =
1.5 M_{\odot}$, dimensionless stellar radius-to-mass ratio $R/M =
5.5$, stellar spin frequency $\nu_* = 582$~Hz, observer's inclination
angle $i = 40^{\rm o}$, $\theta_c = 10^{\rm o}$, $\theta_1 = 40^{\rm
o}$, $\theta_m = 43^{\rm o}$, $\theta_2 = 46^{\rm o}$, $s = 0.004$,
$t_{\rm total} = 6$~s, $t_{\rm rise} = 0.05$~s, $t_{\rm decay} = 8$~s,
$T_{\rm low} = 0.2$~keV, and $T_{\rm high} = 2.6$~keV (see text for
the definitions of the parameters).  The error bars are of $1\sigma$
size.  Vertical dashed lines give the time interval, in which the
radius (and hence the source emission area) does not change much, and
the burning front stalling occurs.}
\end{figure}

\end{document}